\documentclass[showpacs,prl,twocolumn,superscriptaddress,floatfix,amsmath]
   {revtex4} \usepackage{graphicx}
\begin{document}
\title{Effects of thickness on the spin susceptibility of the 2D electron gas}
\author{S. De Palo} \email{depalo@caspur.it} \author{M. Botti}
\email{botti@caspur.it} \affiliation {Dipartimento di Fisica, Universit\`a di
  Roma La Sapienza, P.le Aldo Moro 5, 00185 Roma, Italy} \author{S. Moroni}
\email{moroni@caspur.it} \affiliation{INFM Center for Statistical Mechanics
  and Complexity, Roma, Italy} \affiliation {Dipartimento di Fisica,
  Universit\`a di Roma La Sapienza, P.le Aldo Moro 5, 00185 Roma, Italy}
\author{Gaetano Senatore} \email{senatore@ts.infn.it} \affiliation{INFM DEMOCRITOS
  National Simulation Center, Trieste, Italy} \affiliation{Dipartimento di
  Fisica Teorica, Universit\`a di Trieste, Strada Costiera 11, 34014 Trieste,
  Italy}
\begin{abstract} 
  Using available quantum Monte Carlo predictions for a strictly 2D electron
  gas, we have estimated the spin susceptibility of electrons in actual
  devices taking into account the effect of the finite transverse thickness
  and finding a very good agreement with experiments. A weak disorder, as
  found in very clean devices and/or at densities not too low, just brings
  about a minor enhancement of the susceptibility.
\end{abstract}
\date{}  \maketitle Spin
fluctuations are believed to play an important role in the two dimensional
electron gas (2DEG), near the apparent metal-insulator transition observed at
low temperature in clean devices, with lowering the density\cite{revs}.
Indeed, it has been found that the application of an in--plane magnetic field,
which polarizes the electron spin, suppresses the metallic
conductivity\cite{magne,revs}.  This, together with earlier suggestions that
metallic behavior in 2D should be accompanied by a tendency toward a
ferromagnetic instability\cite{belitz}, has recently prompted a number of
experimental investigations of the spin susceptibility $\chi_s$ of the
2DEG\cite{okam,shas,puda,vitk,tutu,zhu,vaki,shko}, which is generally found to
increase\cite{note} in an appreciable manner when decreasing the density $n$.
Similar behavior is found also on the theoretical side for a strictly 2DEG,
according to the most recent (and most accurate) quantum Monte Carlo (QMC)
results\cite{atta}.  However the susceptibilities measured in different
devices differ among each other and, with one exception\cite{vaki}, do not
agree with the theory\cite{atta}.  Evidently, details of the devices play a
role in determining the properties of the 2DEG and should be accounted for by
the theory, as we shall show below.  In particular in experiments the electron
gas (EG) (i) has a finite transverse thickness, (ii) suffers scattering by a
number of sources (scattering which in fact determines its mobility), and
depending on the system (iii) occupies one or two degenerate valleys. In this
Letter we address points (i) and (ii) for one-valley systems, exploiting the
available QMC data. We find that taking into account the finite thickness of
the specific device quantitatively brings into agreement theory and experiment
and reconciles measurements on different systems, whereas details of
scattering sources play a minor role. Furthermore, we offer some comments on
the effect of (iii) valley degeneracy.

At zero temperature and at given number density $n$, the state of the 2DEG can
be specified by the spin polarization $\zeta=(n_{\uparrow}-n_{\downarrow})/n$.
The spin susceptibility $\chi_s=(\partial \zeta/\partial B)_{B=0}$, which
measures the ratio of the induced spin polarization to an in--plane applied
weak magnetic field $B$, is readily shown to be inversely proportional to the
derivative $(\partial^2 E(\zeta)/\partial \zeta^2)_{\zeta=0}$, involving the
EG internal energy $E(\zeta)$. In fact minimization of the energy per particle
$E(\zeta)+\zeta g_b\mu_B B/2$ with respect to $\zeta$ yields the condition
$E'(\zeta)=-g_b\mu_B B/2$ from which $(\partial \zeta/\partial B)_{B=0}=
-(g_b\mu_B/2)/E''(0)$ immediately follows. An estimate of the spin
susceptibility can be thus obtained from the knowledge of the internal energy
$E(\zeta)$.

The effect of thickness on the 2DEG can be cast, in the simplest
approximation, in terms of a device specific form factor $F(q)$ modifying in
Fourier space the 2D electron-electron interaction $v(q)=2\pi e^2/\varepsilon
q$ into $\tilde{v}(q)=v(q)F(q)$\cite{ando}. Rather than performing new
simulations for each device we have estimated the effects of thickness on
$E(\zeta)$ and hence on $\chi_s$ in a straightforward manner resorting to
perturbation theory. In fact, to the lowest order in $\Delta
v(r)=\tilde{v}(r)-v(r)$, one has for the energy per particle
$E(\zeta)=E_{2D}(\zeta)+\Delta(\zeta)$,
\begin{equation}
\Delta(\zeta)=\frac{n}{2}\int d {\bf r} \Delta v(r) [g_{2D}(\zeta; r)-1],
\label{delta}
\end{equation}
with $E_{2D}(\zeta)$ and $g_{2D}(\zeta; r)$ the known energy\cite{atta} and
pair correlation function\cite{paola}, respectively, of the strictly 2D
electron gas.  The accuracy of the energy estimates obtained in such a manner
has been checked {\it a posteriori } performing selected simulations with the
interaction $\tilde{v}(r)$\cite{sim}.  We have computed the effect of
thickness in two cases, for a GaAs HIGFET\cite{zhu} and for an AlAs quantum
well (QW)\cite{vaki}. In the first case the form factor is\cite{zhang}
\begin{equation}
F(q)= [1+\frac{9}{8} \frac{q}{b} +\frac{3}{8} \frac{q^2}{b^2}]
[1+\frac{q}{b}]^{-3},\label{fgaas}
\end{equation}
with $b^3=48\pi m_b e^2 n^*/\epsilon \hbar^2$ and $n^*=n_d+\frac{11}{32}n$.
Here, $n_d$ is the depletion charge density in the device\cite{stormer},
$m_b=0.067 m_e$ the band electron mass and $\epsilon =12.9 $ the average
background dielectric constant.  For the AlAs QW, on the other hand, the form
factor can be written as\cite{gold}
\begin{equation}
F(q)=\frac{1}{4\pi^2+q^2a^2}\left(3 q a +\frac{8 \pi^2}{q a}-
\frac{32\pi^4}{q^2 a^2}\frac{1-e^{-q a}}{4\pi^2+q^2a^2}\right),
\label{falas}  \end{equation}
with $a=45\AA$ the width of the well\cite{vaki}. Once $F(q)$ is known, it is a
simple matter to evaluate $\Delta(\zeta)$, using Eq.~(\ref{delta}), from which
the enhancement $\chi_s/\chi_P$ of the spin susceptibility $\chi_s$ on its
independent-particle or Pauli value $\chi_P$ is immediately obtained as
$E_0''(0)/E''(0)$, with $E_0(\zeta)=E_F(1+\zeta^2)/2 $ the energy per particle
of non interacting Fermions in 2D and $E_F$ the Fermi energy. Thus
\begin{equation}
\chi_s/\chi_P=E_F/E''(0).\label{chi2}
\end{equation}

\begin{figure}
\includegraphics[width=72mm,angle=-90]{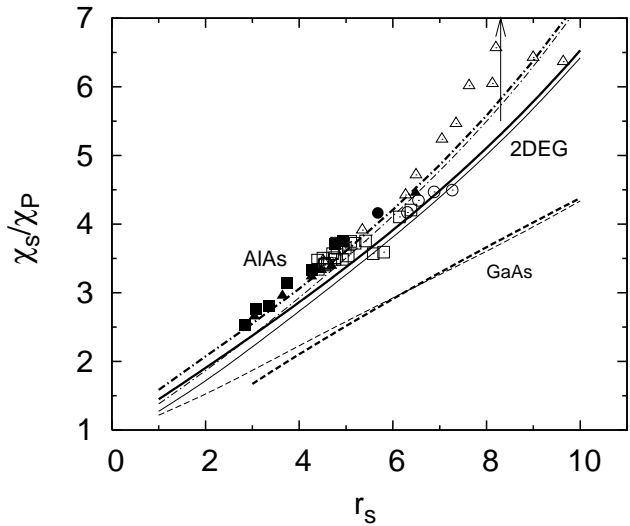}
\caption{Spin susceptibility  in the 2DEG. The thick full curve is the QMC 
  prediction for a strictly 2D system\cite{atta}. Experimental results are
  given by the thick dashed curve for a GaAs HIGFET\cite{zhu} and by various
  symbols, corresponding to different samples, for AlAs QW's\cite{vaki}. The
  thin dashed and full curves add the thickness effect onto the QMC
  prediction\cite{atta} for GaAs and AlAs, respectively.  Finally, the thick
  and thin dot-dashed curves provide the QMC prediction without and with
  inclusion of thickness for AlAs, as obtained from the polarization field
  $B_P$ (see Eq.~(\ref{chip}) in the text). The arrow indicates the location
  of the MIT transition in AlAs.}
\label{chi}
\end{figure}

Our main findings\cite{aps} are summarized in Figure \ref{chi}, which shows a
number of calculations and measurements of $\chi_s/\chi_P$. This quantity is
plotted against the 2D coupling parameter $r_s=U/E_F=1/\sqrt{\pi n}a_B$ to get
rid of uninteresting details of different materials which simply determine the
effective Bohr radius $a_B=\hbar^2\epsilon/m_b e^2$ through the dielectric
constant $\epsilon$ and band mass $m_b$; above, $U=e^2\sqrt{\pi n}/\epsilon$
denotes a rough estimate of the potential energy per particle and
$E_F=\hbar^2\pi n/m_b$.

The QMC prediction for the strictly 2D case\cite{atta} (thick solid line), is
between 30\% and 50\% off the experimental values for GaAs (thick dotted
line). The key result of this paper is that this significant discrepancy is
quantitatively explained as an effect of finite thickness, as clearly shown by
the thin dotted line, obtained via Eqs.~(\ref{delta}, \ref{fgaas}) and
(\ref{chi2}). This conclusion is further strengthened by our explicit estimate
of the effect of weak disorder, due to background doping in the GaAs
HIGFET\cite{zhu}, which turns out to be negligible (see below).  We emphasize
that the parameters entering the form factor of Eq.~(\ref{fgaas}) reflect our
knowledge of the real sample\cite{stormer}, and they are not adjusted to
achieve a particular value of the spin susceptibility.

\begin{figure}
\includegraphics[width=60mm,angle=-90]{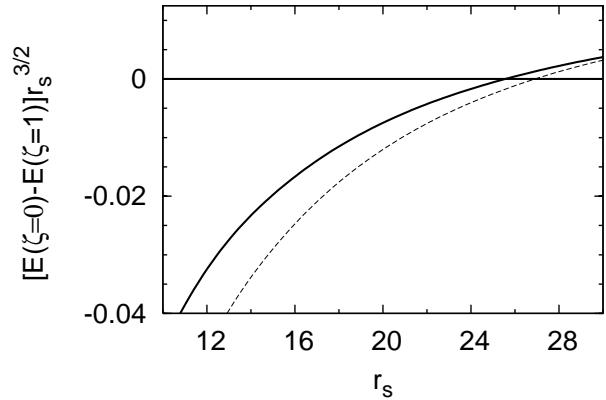}
\caption{Difference between the QMC energies of the unpolarized and  polarized
  phases for the strictly 2DEG (solid line) and with the thickness of the GaAs
  HIGFET (dashed line).}
\label{diag}
\end{figure}

In view of the interest for a possible ferromagnetic instability in
low--density 2D electron systems, the question arises as to whether thickness,
which noticeably changes the spin susceptibility, also alters the stability
range of the polarized fluid ($26\lesssim r_s\lesssim 35$) predicted by QMC in
the strictly 2D case\cite{atta}.  Figure \ref{diag} shows that this stability
window is only slightly shrunk (by $\lesssim 2$ in $r_s$) and the
ferromagnetic instability pushed at slightly higher $r_s$.

For the AlAs device the situation is somewhat different.  In order to engineer
a one--valley 2DEG with isotropic mass, the electrons are confined in a very
narrow QW, thus reducing the importance of finite thickness effects, but also
boosting the possible influence of well width fluctuations\cite{vaki}. The
spin susceptibility, measured with either the tilted field or the polarization
field methods (filled and empty symbols in Fig. \ref{chi}, respectively),
turns out to be fairly close to the strictly 2D QMC value.  As expected, the
tiny width of the QW does not affect this value significantly, as shown by the
thick and the thin full curves in Fig.~\ref{chi}. Therefore the small
discrepancy between the QMC prediction (with or without finite thickness) and
the experimental result points to a possible role of disorder.

Before discussing our estimates of the effect of disorder, however, it is
worth to briefly comment on the two different techniques used to measure
$\chi_s/\chi_P$. Within the Landau--Fermi liquid theory, the susceptibility
enhancement can be expressed in terms of the quasiparticle parameters $g^*$
and $m^*$ as $\chi_s/\chi_P=g^*m^*/g_bm_b$, where $m_b$ and $g_b$ are the mass
and g-factor entering the hamiltonian describing the interacting electrons,
which for electrons in a device coincide with the band mass and g-factor. One
of the experimental techniques employed to estimate $\chi_s$ is the tilted
field method of Fang and Stiles\cite{fang}, which allows the determination of
$g^*m^*$ from the analysis of the minima in the Shubnikov-de Haas
oscillations. The experimental results in Fig.~\ref{chi} for the HIGFET and
part of those for the QW's (full symbols) were obtained with this technique.
An alternative manner to extract $g^*m^*$ from experiments has been suggested
by Okamoto\cite{okam}. If the interacting electrons can be replaced by
independent particles with effective parameters ($ m^*$ and $g^*$), then the
(in--plane) magnetic field necessary to induce full spin polarization
satisfies $\mu_B g^* B_P=2 E_F^*=\hbar^2 2 \pi n/m^*$, which gives
$\chi_s/\chi_P=2E_F/g_bm_bB_P$. However, the (in--plane) polarization field
must also satisfy the exact condition $g_b\mu_B B_P=2E'(1)$, which combined
with the above yields
\begin{equation}
\chi_s/\chi_P=E_F/E'(1).\label{chip}
\end{equation}
The experimental results for the AlAs QW's obtained with both techniques are
consistent with each other, due to the spread in the data.  On the other hand,
this is not the case with QMC\cite{atta} for which Eq.~(\ref{chip}) yields an
appreciable overestimate of the susceptibility enhancement. While the correct
definition of the spin susceptibility is the one of Eq.~(\ref{chi2}), in
comparing measured and calculated values of $\chi_s/\chi_P$ it is appropriate
to refer to theoretical estimates consistent with the adopted experimental
determination. In particular for low density, where only polarization field
data are available (say $r_s$ larger than about 6), the theoretical value to
be considered is the dash-dotted lines, calculated using Eq.~(\ref{chip}).

We now turn to the discussion of the disorder effects.  A realistic
description of these devices needs the inclusion of the different scattering
sources that determine the mobility at zero temperature. The GaAs
HIGFET\cite{zhu} is a very clean device with no intentional doping and with a
concentration of background impurities of $\approx 5 \cdot 10^{12} cm^{-3} $,
which is indeed a very low value\cite{modeldis}.  Such a concentration has
obtained through a best fit of the measured mobility\cite{zhuthesis}, as a
function of the electron density, computed using Born
approximation\cite{response1}.  We estimate the gross effect of such weak
disorder on the spin susceptibility by means of perturbation theory,
describing the disorder in terms of an external one-body potential $u({\bf
  r})$, coupling to the electron density, with known first and second moment
ensemble averages\cite{disorder}. As usual we assume a vanishing first moment.
The first non vanishing contribution to the energy reads:
\begin{equation}
\Delta_d(\zeta)=\frac{1}{2n} \int \frac{d{\bf q}}{(2 \pi)^2} \chi_{n,n}(q,
\zeta)\frac{\langle u({\bf q})^2\rangle}{A}, \label{ddis}
\end{equation}
with $u({\bf q})$ the 2D Fourier transform of the one-body potential and
$\langle \cdots\rangle$ denoting the ensemble average over disorder
configurations, per unit area. Above, $ \chi_{n,n}(q, \zeta)$ is the
density--density response function (in Fourier space) at polarization $\zeta$
for the strictly 2DEG\cite{response1}.

\begin{figure}
\includegraphics[width=72mm,angle=-90]{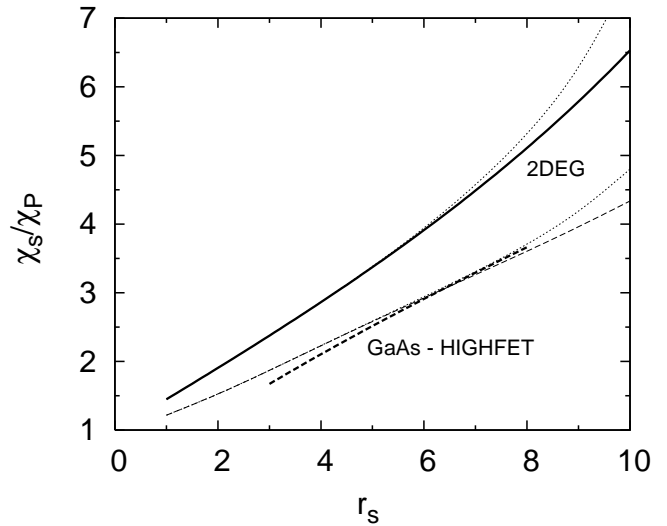}
\caption{Effect of a weak disorder appropriate to the  GaAs HIGHFET on the 
  spin susceptibility.  Upper and lower dotted curves are for the strictly
  2DEG and for a 2DEG with the GaAs HIGHFET thickness. Other curves as in
  Fig.~\ref{chi}.}
\label{chidis}
\end{figure}

In Fig.~\ref{chidis} we show our results for the GaAs HIGFET, using
Eq.~(\ref{ddis}) to estimate the effect of disorder in a density range in
which $\Delta_d(\zeta)$ is much smaller (less than 1\%) than the unperturbed
2DEG energy .  The effect of disorder is to enhance the spin susceptibility,
contrary to that of transverse thickness.  However, in the experimental
density range ($3\lesssim r_s\lesssim 8$) such an effect is negligible.

The mobility in AlAs QW's is roughly two orders of magnitude smaller than in
the GaAs HIGFET, as the scattering sources are more effective. We were able to
reproduce the measured mobilities\cite{vaki}, using Born
approximation\cite{response1,alas}, only up to $r_s\lesssim 4$. In this
density range the additional enhancement of $\chi_s$ due to disorder remains
$\lesssim 10\%$, yielding a good agreement between the prediction of
Eq.~(\ref{chi2}), with $E(\zeta)=E_{2D}(\zeta)+\Delta(\zeta)+\Delta_d(\zeta)$,
and the available results of the tilted field experiments. Our findings
suggest that, far from the metal-insulator transition, occurring at
$r_s^c=8.3$ in AlAs QW's and at $r_s^c=12.4$ in the GaAs HIGFET, disorder
effects yield a small enhancement of $\chi_s$, which is either negligible as
in GaAs or helps in reducing the small residual discrepancy between QMC and
experiments.  However, to obtain indications valid at larger $r_s$ and/or
stronger disorder, an approach that takes into account disorder and
electron-electron interaction on the same footing is required.

We finally comment on the two--valley electron systems, realized in Si
MOSFETs\cite{okam,shas,puda,vitk} and in wide AlAs QW\cite{shko}. In
particular Shkolnikov et al\cite{shko}, tuning the valley population, have
shown that valley degeneracy brings about a depression of the spin
susceptibility, in sharp qualitative contrast with the enhancement predicted
by Hartree-Fock theory. A first estimate of the spin susceptibility of a
two--valley symmetric EG can be simply obtained from previous QMC studies of
the energy of four--\cite{conti} and two--component\cite{rapi} electrons
performed with the same level of accuracy, assuming a quadratic dispersion of
the energy $E(\zeta)$ with $\zeta$\cite{note2}. Such an estimate clearly shows
the qualitative effect observed in Ref. \onlinecite{shko}.  A detailed
comparison with either the Si MOSFET\cite{okam,shas,puda,vitk} or the
anisotropic--mass AlAs QW\cite{shko} devices along the lines of the present
calculation would require QMC input which is presently not available. We are
currently performing extensive simulations of the strictly 2D two--valley
system with finite polarization to obtain an accurate theoretical prediction
of $\chi_s$ in such a system.

In conclusion we have shown that a realistic description of actual devices,
starting from the 2DEG model and including specific features of the systems,
enables us to reproduce the spin susceptibility without any adjustable
parameters. In GaAs HIGFET\cite{zhu} the thickness plays a crucial role, while
the weak disorder provides only a negligibly small enhancement.  In the AlAs
QW's case\cite{vaki} the 2DEG spin susceptibility is directly comparable with
experiments. A residual small discrepancy is likely due to the influence of
QW's width fluctuations.


\begin{thebibliography}{99}
\bibitem{revs}E. Abrahams, S. V. Kravchenko and M. P. Sarachik,
Rev. Mod. Phys.{\bf 73}, 251 (2001); S. V. Kravchenko and M. P. Sarachik,
Rep. Prog. Phys. {\bf 67}, 1 (2004); and reference therein.
\bibitem{magne}V. T. Dolgopolov, G. V. Kravchenko, A. A. Shashkin, and
S. V. Kravchenko, Phys. Rev. {\bf B 46}, 13303 (1992).
\bibitem{belitz} C. Castellani, C. Di Castro, and P.A. Lee, Phys. Rev. B {\bf
57}, R9381, 1998; see, also, C. Castellani, C. Di Castro, P. A. Lee, M. Ma,
S. Sorella, and E. Tabet, Phys. Rev. B {\bf 33}, 6169 (1986); D. Belitz, and
T.R. Kirkpatrick, Rev. Mod. Phys. {\bf 66}, 261 (1994).
\bibitem{okam} T. Okamoto, K. Hosoya, S. Kawaji, and A. Yagi,
Phys. Rev. Lett. {\bf 82}, 3875 (1999).
\bibitem{shas} A. A. Shashkin, S. V. Kravchenko, V. T. Dolgopolov, and
T. M. Klapwijk. Phys. Rev. Lett. {\bf 87}, 086801 (2001).
\bibitem{puda}V. M. Pudalov, M. E. Gershenson, H. Kojima, N. Butch,
E. M. Dizhur, G. Brunthaler, A. Prinz and G. Bauer, Phys. Rev. Lett. {\bf 88
}, 196404 (2002)
\bibitem{vitk} S.A. Vitkalov, M.P Sarachik, and T.M. Klapwijk, Phys. Rev. B
{\bf 65}, 201106(R) (2002).
\bibitem{tutu}E. Tutuc, S. Melinte, and M Shayegan, Phys. Rev. Lett. {\bf 88},
36805 (2002); E. Tutuc et al, Phys. Rev. B {\bf 67}, 241309 (R) (2003).
\bibitem{zhu} J. Zhu, H.L. Stormer, L. N. Pfeiffer, K. W. Baldwin and
K.W. West, Phys. Rev. Lett. {\bf 90}, 56805 (2003).
\bibitem{vaki}K. Vakili, Y.P. Shkolnikov, E. Tutuc, E.P. De Poortere, and
M. Shayegan, Phys. Rev. Lett. {\bf 92}, 226401 (2004). 
\bibitem{shko} Y. P. Shkolnikov, K. Vakili, E. P. De Poortere, and
M. Shayegan, Phys. Rev. Lett. {\bf 92}, 246804 (2004).
\bibitem{note} The results of Ref. \onlinecite{tutu} show a decrease of
$\chi_s$ with $n$, but see Refs. \onlinecite{tutu,zhu} for an explanation.
\bibitem{atta} C. Attaccalite, S. Moroni, P. Gori-Giorgi, and G B. Bachelet,
Phys. Rev. Lett. {\bf 88}, 256601 (2002)
\bibitem{ando} T. Ando, A. B. Fowler, and Frank Stern,.  Rev. Mod. Phys. {\bf
54}, 437 (1982)
\bibitem{paola} P. Gori-Giorgi, S. Moroni and G.B. Bachelet, Phys. Rev. {\bf B
70}, 115102 (2004).
\bibitem{sim} We have performed simulations with interaction $\tilde{v}(r)$,
with parameters corresponding to the HIGFET\cite{zhu}, at $r_s=5$ and
$\zeta=0,1$, and determined $\Delta=E(\zeta)-E_{2D}(\zeta)$; we find a
fractional difference from the first order estimate of Eq.~(\ref{delta}) of
only a few percent.
\bibitem{zhang} F.C. Zhang and S. Das Sarma, Phys. Rev. {\bf B 33}, 2903R
(1986); and references there in.
\bibitem{stormer} For the considered HIGFET the depletion is negligible
(H. Stormer private communication) and we set $n_d=0$.
\bibitem{gold} A. Gold, Phys. Rev. B. {\bf 35},  723 (1987).
\bibitem{aps}A brief account of these results has been given  at the
  2004 APS March  Meeting,  presentation W11.015 [S. De Palo, G. Senatore,
  M. Botti, and S. Moroni,   Bull. Amer. Phys. Soc. {\bf 49}, 1303 (2004)].
\bibitem{fang} F. F. Fang and P. J. Stiles. Phys. Rev. {\bf 174}, 823 (1968)
\bibitem{modeldis} We refer the reader to \cite{gold2} for the description of
disorder in GaAs Heterostructures and to \cite{gold} for QW's, noting that
Eq.~(3c) of \cite{gold} has misprints.
\bibitem{zhuthesis} Jun Zhu, Ph.D. Thesis, Columbia University (2003).
\bibitem{response1} S. De Palo and G. Senatore, unpublished.
\bibitem{disorder}Evidently, one has to calculate the EG properties
for a given configuration of disorder and then average the properties on 
disorder configurations (ensemble average).
\bibitem{alas}We include disorder parameters for the real device(K. Vakili
  private communication), adjusting only the parameters for the QW width
  fluctuations (roughness scattering) , to obtain agreement with the
  measured mobility.
\bibitem{conti} S. Conti and G. Senatore, Europhys. Lett. {\bf
36}, 695 (1996).
\bibitem{rapi} F. Rapisarda and G. Senatore, Aust. J. Phys. {\bf 49}, 161
(1996).
\bibitem{note2}
Evidently, if $E_{1v}(\zeta)$ and $E_{2v}(\zeta)$ are the energy of the
two--component (spin degeneracy) and four--component (spin and valley
degeneracy) systems, we set $E_{2v}(\zeta)\approx E_{2v}(0)+
(E_{2v}(1)-E_{2v}(0))\zeta^2$ and exploit the fact that  for symmetric 
valleys  $E_{2v}(1)=E_{1v}(0)$.
\bibitem{gold2}See, e.g., A. Gold, Appl. Phys. Lett. {\bf 54}, 2100 (1989).
\end{thebibliography}
\end{document}